\documentclass
[aps,prl,amsfonts,amssymb,twocolumn,amsmath,preprintnumbers,nofootinbib,floatfix,superscriptaddress,longbibliography]{revtex4-1}%
\usepackage[dvips]{graphics}
\usepackage{graphicx}
\usepackage{bm}
\usepackage{amsmath, bm}
\usepackage{amsfonts}
\usepackage{amssymb}
\usepackage{xcolor}
\usepackage{subfigure}
\usepackage{tabularx}
\usepackage{multirow}
\usepackage{hyperref,hypcap}
\usepackage{braket}
\usepackage{commath}%
\usepackage{lineno}

\makeatletter
\def\maketitle{
	\@author@finish
	\title@column\titleblock@produce
	\suppressfloats[t]}
\makeatother

\providecommand{\U}[1]{\protect\rule{.1in}{.1in}}

\newcommand{\bk}{\mathbf{k}}
\newcommand{\bK}{\mathbf{K}}
\newcommand{\br}{\mathbf{r}}
\newcommand{\bG}{\mathbf{G}}
\newcommand{\bd}{\mathbf{d}}
\newcommand{\bs}{\mathbf{s}}

\begin{document}

\title{Time-Reversal Even Charge Hall Effect from Twisted Interface Coupling}
\author{Dawei Zhai}
\thanks{These authors contributed equally to this work.}
\author{Cong Chen}
\thanks{These authors contributed equally to this work.}
\author{Cong Xiao}
\email{congxiao@hku.hk}
\author{Wang Yao}
\email{wangyao@hku.hk}
\affiliation{Department of Physics, The University of Hong Kong, Hong Kong, China}
\affiliation{HKU-UCAS Joint Institute of Theoretical and Computational Physics at Hong Kong, China}

\maketitle

\noindent{\bf{Abstract}}\\
{
Under time-reversal symmetry, a linear charge Hall response is usually deemed to be forbidden by the Onsager relation. In this work, we discover a scenario for realizing a time-reversal even linear charge Hall effect in a non-isolated two-dimensional crystal allowed by time reversal symmetry. The restriction by Onsager relation is lifted by interfacial coupling with an adjacent layer, where the overall chiral symmetry requirement is fulfilled by a twisted stacking. We reveal the underlying band geometric quantity as the momentum-space vorticity of layer current. The effect is demonstrated in twisted bilayer graphene and twisted homobilayer transition metal dichalcogenides with a wide range of twist angles, which exhibit giant Hall ratios under experimentally practical conditions, with gate voltage controlled on-off switch. This work reveals intriguing Hall physics in chiral structures, and opens up a research direction of layertronics that exploits the quantum nature of layer degree of freedom to uncover exciting effects.
}


\noindent{\bf{Introduction}}\\
Hall effect, by its rigorous definition, refers to a transverse charge current $\mathbf{j}_{\text{H}%
}=\boldsymbol{\sigma}_{\text{H}}\times\mathbf{E}$,
with a unidirectional or chiral nature characterized by the Hall
conductivity pseudovector $\boldsymbol{\sigma}_{\text{H}}$~\cite{Nagaosa2010}. In principle, $\boldsymbol{\sigma}_{\text{H}}$ can have two parts that are odd and even respectively under time
reversal (TR). The TR-odd part, such as the ordinary Hall effect induced by
Lorentz force and the anomalous Hall effect induced by the momentum space
Berry curvature (Fig.~\ref{Fig:Scheme}a), is obviously forbidden in TR
symmetric systems. The TR-even part is also forbidden under TR symmetry, with a more delicate origin in the Onsager relation of electrical conductivity~\cite{Nagaosa2010,Onsager1}. Therefore, the linear-response charge Hall transport has been observed only in
materials with TR breaking by magnetic field or magnetic order~\cite{Nagaosa2010,Weng2015,Qi2016}.

\begin{figure}[h]
	\includegraphics[width=8.8cm]{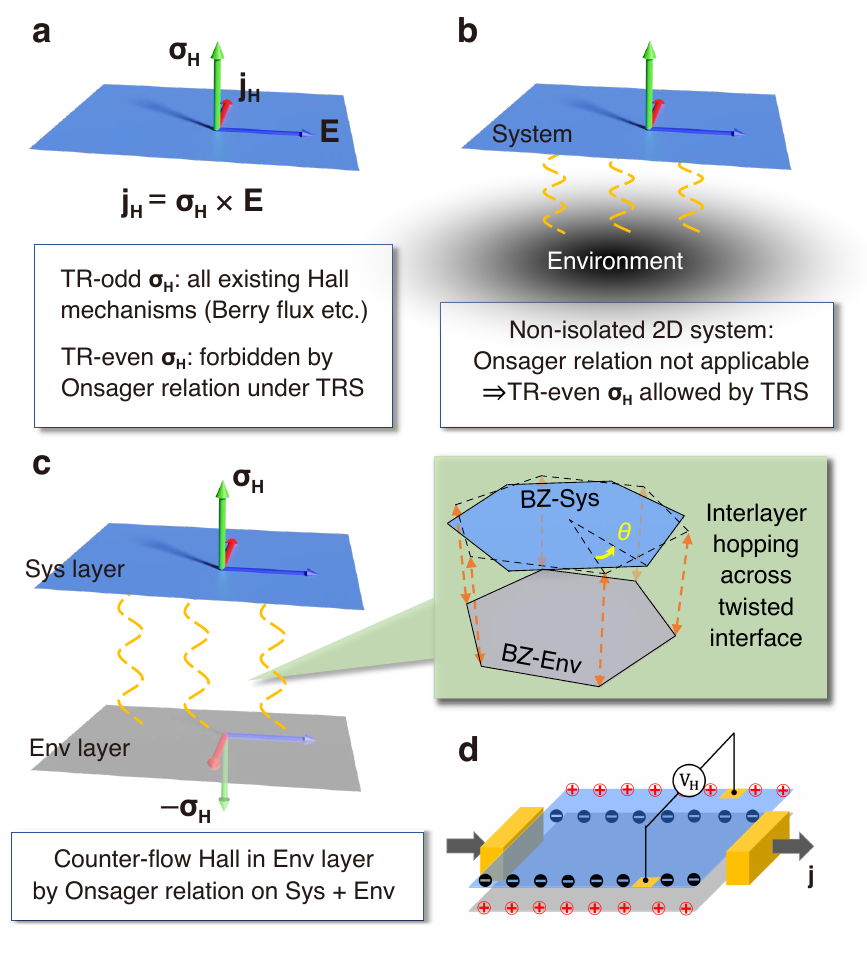} 
	\caption{\textbf{Linear charge Hall effect under time reversal (TR) symmetry in a non-isolated system.} \textbf{a} In the Hall
		conductivity $\boldsymbol{\sigma_{\text{H}}}$, the TR odd part by definition requires TR symmetry breaking, while the TR even part vanishes by Onsager relation. The red, green, and blue arrows denote the three vectors in $\mathbf{j}_{\text{H}}=\boldsymbol{\sigma}_{\text{H}}\times\mathbf{E}$.
		\textbf{b} Constraint by Onsager relation can be lifted in a non-isolated 2D system (blue surface) by coupling to an environment (black shading). \textbf{c} TR-even Hall current from twisted interfacial coupling with an environmental layer (grey surface), whereas a counterflow Hall current is expected in the latter, by Onsager relation on the whole structure: system (Sys) + environment (Env). Green shaded area denotes the interlayer hopping between the Brillouin zone (BZ) with twist angle $\theta$. \textbf{d} The TR-even Hall voltage ($V_H$) due to charge accumulation at the sample edges (red and black $+/-$) can be detected with a layer-resolved measurement. Black arrows denote source and drain current $\mathbf{j}$.}%
	\label{Fig:Scheme}%
\end{figure}

For a non-isolated system, however, the Onsager relation on electrical conductivity is not necessarily applicable~\cite{Onsager1}, depending on the form of its interplay with the environment. This in fact leaves room for the TR-even contribution to $\boldsymbol{\sigma}_{\text{H}}$ in the system, and hence the possibility of having charge Hall effect under TR symmetry (Fig.~\ref{Fig:Scheme}b).
If the system is geometrically separated from the environment
for the Hall voltage to be measurable, the TR-even Hall effect can have real impact, besides being fundamentally intriguing.

A platform to explore the above scenario is naturally provided in twisted van der Waals (vdW) layered structures~\cite{moireReviewEvaMacDonaldNatMater2020,moireReviewNatPhysBalents2020,moireReviewRubioNatPhys2021,moireReviewExptFolksNatRevMat2021,moireReviewJeanieLauNature2022,moireexcitonreviewNature2021,moireexcitonreviewNatRevMater2022},
where a two-dimensional (2D) crystal is separated by the vdW gap from adjacent layers constituting its environment (Fig.~\ref{Fig:Scheme}c).
With the demonstrated capability to access electrical conduction in individual layers of the vdW structures~\cite{BilayerHallGeimNatPhys2012,BilayerHallPhilipKimNatPhys2017,BilayerHallPabloNatNano2017,BilayerHallPhilipKimNatPhys2019,BilayerHallCoryDeanScience2022,BilayerHallPhilipKimPRL2017}, this definition of system and environment becomes physical. The restriction by Onsager relation on the system layer's conduction is lifted by quantum tunneling of electrons across the vdW gap. Without magnetic field or magnetic order, nevertheless, chiral structural symmetry is required instead to comply with the chiral nature of Hall current. This is fulfilled in a twisted stacking that breaks inversion and all mirror symmetries~\cite{TobiasPRB2020}. The intriguing TR-even Hall effect, nonetheless, remains unexplored.

In this work, we demonstrate the TR-even charge Hall effect in a twisted double layer with TR symmetry. Hall transport and
charge accumulations at edges are made possible in an individual layer.
Meanwhile, the Onsager relation for the whole double-layer geometry demands an opposite Hall
flow in the environment layer (Fig.~\ref{Fig:Scheme}c).
We find that the TR-even Hall response here is rooted in a band geometric quantity -- the momentum space
vorticity of layer current -- that emerges from the interlayer hybridization of
electronic states under TR and chiral symmetry.
Our symmetry analyses show that the effect is
characteristic of general chiral bilayers with Fermi surface,
which we quantitatively demonstrate for the exemplary systems of twisted bilayer graphene (tBG) and twisted homobilayer transition metal dichalcogenides (tTMDs) with a wide range of twist angles. Within experimentally feasible range of carrier doping~\cite{Zhang2015gate,Nam2017}%
, we find pronounced Hall responses accompanied by giant Hall ratios (e.g., $\mathcal{O}(1)$ in tBG), with sign and magnitude controlled by the twist
angle. The effects in tTMDs also feature good on-off switchability by
gate voltage that promises device applications.\\

\noindent{\bf{Results}}\\
\textbf{Symmetry characters.}
The charge Hall counterflow in system and environment layers leads to accumulation of interlayer charge dipoles (i.e., opposite charges in the two layers) near the transverse edges (Fig.~\ref{Fig:Scheme}d). Accordingly, it can be holistically viewed as the Hall transport of the charge dipole moment. The dipole Hall current measures the difference of the Hall flows in the two layers, i.e., $\textbf{j}^{d}_{\text{H}}=\textbf{j}^{\text{sys}}_{\text{H}}-\textbf{j}^{\text{env}}_{\text{H}}$. Onsager relation forbids a net charge Hall current counting both layers, $\textbf{j}^{\text{sys}}_{\text{H}} + \textbf{j}^{\text{env}}_{\text{H}} = 0$, therefore,
\begin{equation}
\sigma_{yx}^{\text{sys}}=-\sigma_{yx}^{\text{env}}=\sigma_{yx}^{d}/2, \label{relation}
\end{equation}
where $\sigma_{yx}^{\text{sys}/\text{env}}$ quantifies the charge Hall effect
in the system/environment layer (Fig.~\ref{Fig:Scheme}c).
Such a perspective is particularly useful for identifying the symmetry requirements for the appearance of the TR-even Hall effect as elaborated in the following.

The dipole current generated at the linear
order of a driving electric field \textbf{E} is given by $j_{a}%
^{d}=\sigma_{ab}^{d}E_{b}$, where the Einstein summation convention is adopted
for in-plane Cartesian Coordinates $a$ and $b$, and $j_{a}^{d}$ is the current
along the $a$ direction of the out-of-plane interlayer dipole moment. Since
the dipole current is odd under TR while the electric field is even,
in nonmagnetic metallic states, the effect can only stem from nonequilibrium kinetics of electrons around the Fermi surface, and the resulting $\sigma_{ab}^{d}$ is a TR-even tensor. The dipole Hall conductivity,
which is antisymmetric with respect to the directions of the in-plane driving
field and response current, is dual to $\sigma_{\text{H}}^{d}=(\sigma_{yx}^{d}-\sigma_{xy}^{d})/2$ that transforms as a pseudoscalar (the $zz$ component of a TR-even pseudotensor). Namely, it remains unchanged under
rotation, but changes sign under space inversion, mirror reflection, and
roto-reflection. Such a dipole Hall effect under TR symmetry is therefore allowed provided that
the bilayer crystal structure is chiral.

This chiral symmetry requirement is fulfilled in the twisted bilayer
vdW structures, such as tBG and tTMDs. These most studied
structures, which are also our foci, are based on honeycomb lattices
and preserve the threefold rotation symmetry in the $z$ direction.
This symmetry forbids the off-diagonal components of the symmetric part of
$\sigma_{ab}^{d}$ with respect to $a$ and $b$, i.e., $\sigma_{yx}^{d} + \sigma_{xy}^{d} = 0$. Therefore, according to Eq.~(\ref{relation}), the TR-even charge Hall current in the
system layer is quantified by
\begin{equation}
	\sigma_{\text{H}}^{\text{sys}}=\sigma_{\text{H}}^{d}/2=\sigma_{yx}^{\text{sys}}=-\sigma_{xy}^{\text{sys}}.
\end{equation}

We point out that the direction of Hall current can be reversed with an opposite twist direction (Fig.~\ref{Fig:TBG_theta}d). This can be easily understood by noticing that structures obtained with twist angle $\theta$ and $-\theta$ are mirror images of each other, whereas the mirror operation flips the sign of the Hall current in each layer.

\begin{figure*}[t]
	\centering
	\includegraphics[width=14cm]{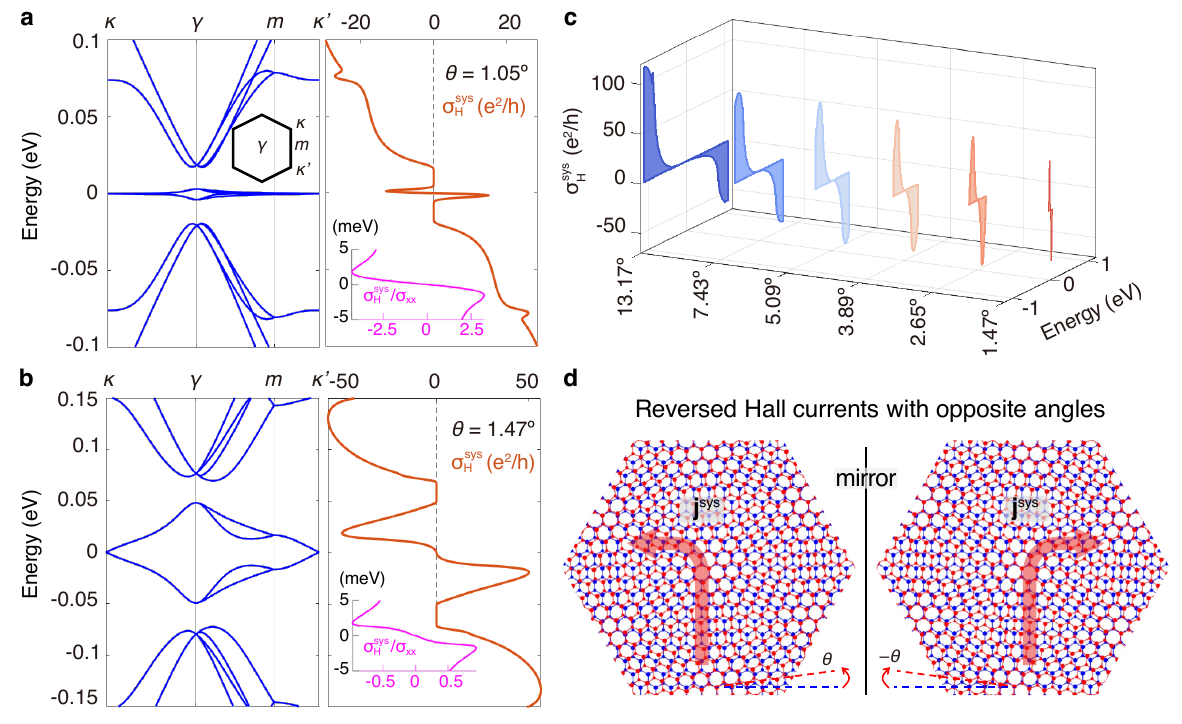} 
	\caption{\textbf{Band structure and TR-even Hall conductivity for tBG at different $\theta$}. \textbf{a} $\theta=1.05^{\circ}$. \textbf{b} $\theta=1.47^{\circ}$. Insets: Hall ratio of Hall conductivity $\sigma_{\text{H}}^{\text{sys}}$ to the longitudinal conductivity $\sigma_{xx}$ when the Fermi level is within the central bands. \textbf{c}
		Evolution of the two central peaks of $\sigma_{\text{H}}^{\text{sys}}$ with $\theta$.
		Here $\sigma_{\text{H}}^{\text{sys}}$ should be
		multiplied by a factor of 2 accounting for spin degeneracy. \textbf{d} Schematics of reversed Hall currents in the system layer $\mathbf{j}^{\text{sys}}$ by twisting in opposite directions, where the moir\'e lattices are mirror images of each other. In the
		calculations here and hereafter we take $\tau=1$ ps.}%
	\label{Fig:TBG_theta}%
\end{figure*}


\textbf{General theory: k-space vorticity of layer current.}
The current density in the system and environment layer is given by the integral of the layer resolved current
$e\mathbf
v_{n}^{\text{sys}/\text{env}}(\mathbf{k})$ carried by each electron weighted by the
distribution function $f_{n}(\mathbf{k})$:%
\begin{equation}
\mathbf{j}^{\text{sys}/\text{env}}=e\sum_{n}\int\frac{d^{2}\mathbf{k}}{(2\pi)^{2}}f_{n}%
(\mathbf{k})\mathbf{v}_{n}^{\text{sys}/\text{env}}(\mathbf{k}), \label{S}%
\end{equation}
where $n$ and $\hbar\mathbf{k}$ are the band index and crystal momentum.
$\mathbf{ v}_{n}^{\text{sys}/\text{env}}(\mathbf{k}) = \langle u_{n}(\mathbf{k}%
)|   \frac{1}{2}\{\mathbf{\hat{v}},\hat{P}^{\text{sys}/\text{env}}\}  |u_{n}(\mathbf{k})\rangle$, with
$\hat{P}^{\text{sys}}=(1 + \hat{l}^{z})/2$ and $\hat{P}^{\text{env}}=(1- \hat{l}^{z})/2$ being respectively the
projection operator onto the system and environment layer, and
$\hat{l}^{z}=\text{diag}(1,-1)$ operating in the layer
index subspace~\cite{Xu2014}. Because of the TR symmetry, $\mathbf{v}_{n}%
^{\text{sys}/\text{env}}(\mathbf{k})=-\mathbf{v}_{n}^{\text{sys}/\text{env}}(-\mathbf{k})$, hence a nonzero
layer current requires a distribution function in $k$-space that breaks the
occupation symmetry at $\mathbf{k}$ and $-\mathbf{k}$. Such an
off-equilibrium distribution can be driven by an electric field and described
by the Boltzmann transport equation. Within the simplest constant relaxation
time approximation, the deviation from the equilibrium Fermi distribution
$f_{0}\equiv f_{0}(\varepsilon_{n})$ is of a dipole structure in $k$-space:
$f_{n}-f_{0}=-\frac{e}{\hbar}\tau\mathbf{E} \cdot\partial_{\mathbf{k}}f_{0}$,
with $\varepsilon_{n}$ being the band energy and $\tau$ the transport
relaxation time. This approximation is usually taken so that the specific
content of disorder often unknown does not pose a difficulty and that the
band origin of the effect can be manifested~\cite{Xie2021,Fu2015,Lai2021}.

We focus on the TR-even charge Hall response in the system layer in the following. The Hall conductivity reads%
\begin{equation}
\sigma_{\text{H}}^{\text{sys}}=\frac{e^{2}}{\hbar}%
\tau\mathcal{V},
\end{equation}
where%
\begin{equation}
\mathcal{V}=-\frac{\hbar}{2}\sum_{n}\int\frac{d^{2}\mathbf{k}}{(2\pi)^{2}%
}f_{0}^{\prime}\left[  \mathbf{v}_{n}(\mathbf{k})\times\mathbf v_{n}^{\text{sys}}%
(\mathbf{k})\right]  _{z}\label{FS}%
\end{equation}
is intrinsic to the band structure, has the dimension of frequency, and is
indeed a TR-even pseudoscalar conforming to the symmetry
analysis. Here $\mathbf{v}_{n} (\mathbf{k}) = \langle u_{n}(\mathbf{k}%
)|  \mathbf{\hat{v}}  |u_{n}(\mathbf{k})\rangle = \mathbf{v}_{n}^{\text{sys}} (\mathbf{k}) +\mathbf{v}_{n}^{\text{env}} (\mathbf{k}) $, and
$f_{0}^{\prime}=\partial f_{0}/\partial\varepsilon_{n}$ implies that the TR-even
Hall effect is a Fermi surface property. If interlayer coupling is absent,
one has $\mathbf{v}_{n}^{\text{sys}}(\mathbf{k})=\mathbf{v}_{n}(\mathbf{k})[1+
\l_{n}^{z}(\mathbf{k})]/2$, hence no TR-even Hall effect. Similarly, the Hall conductivity of the environment layer can be obtained by replacing $\mathbf{v}_{n}^{\text{sys}}(\mathbf{k})$ in $\mathcal{V}$ with $\mathbf{v}_{n}^{\text{env}}(\mathbf{k})$. It is clear that $\sigma_{\text{H}}^{\text{env}}=-\sigma_{\text{H}}^{\text{sys}}$ and
the total charge Hall current of a bilayer geometry is indeed vanishing. The
formal theory thus confirms the conclusions of the foregoing symmetry arguments.

Now we show that the TR-even Hall effect has a band origin in the \textit{k}-space vorticity of the layer current. Via integration by parts, Eq.~(\ref{FS}) is
recast into%
\begin{equation}
\mathcal{V}=\sum_{n}\int\frac{d^{2}\mathbf{k}}{(2\pi)^{2}}f_{0}\,%
\mathcal{\omega}_{n}\left(  \mathbf{k}\right)  , \label{sea}%
\end{equation}
which measures the \textit{k}-space vorticity of the layer current
$\mathcal{\omega}_{n}\left(  \mathbf{k}\right)  =\frac{1}{2}%
[\bm\partial_{\mathbf{k}}\times\mathbf{v}_{n}^{\text{sys}}(\mathbf{k})]_{z}$
integrated over the occupied states. As the integral of this vorticity over
any full band vanishes, only $\mathcal{\omega}_{n}\left(  \mathbf{k}%
\right)  $ of partially occupied bands contribute to a net TR-even Hall effect.
The layer current vorticity can be expressed in an enlightening form
\begin{equation}
\mathcal{\omega}_{n}\left(  \mathbf{k}\right)  =\hbar\operatorname{Re}%
\sum_{n_{1}\neq n}\frac{[\mathbf{v}_{nn_{1}}\left(  \mathbf{k}%
\right)  \times\mathbf{v}_{n_{1}n}^{\text{sys}}\left(  \mathbf{k}\right)
]_{z}}{\varepsilon_{n}\left(  \mathbf{k}\right)  -\varepsilon_{n_{1}%
}\left(  \mathbf{k}\right)  }, \label{quantity}%
\end{equation}
where the numerator involves interband matrix elements of total velocity and layer
velocity operators.
Under the TR operation, interband quantities are transformed into their complex conjugates and $\varepsilon_{n}(\mathbf{k})$ is even, with which one finds that $\omega_{n}(\mathbf{k})$ is also TR-even after taking the real part.
This expression of $\mathcal{\omega}_{n}\left(
\mathbf{k}\right)  $ shares a striking similarity with the well-known band geometric
quantity \textit{k}-space Berry curvature~\cite{Xiao2010}: The former
becomes the latter if $\mathbf{v}_{n_{1}n}^{\text{sys}}$ is replaced by the
$k$-space interband Berry connection $\mathbf{A}_{nn_{1}}=\langle
u_{n}|i\partial_{\mathbf{k}}|u_{n_{1}}\rangle$.
As such, the TR-even Hall effect,
despite being described by the Boltzman transport theory, encodes the
information of interband coherence, which has up to now mostly connected to
intrinsic transport effects induced by various Berry-phase
effects~\cite{Nagaosa2010,Sinova2015,Xu2016valleytronics,Xiao2010}, and hence should be
enhanced when the Fermi level is located around band near-degeneracy regions.
Despite the similarity, we stress that the layer current vorticity is fundamentally different from the \textit{k}-space Berry curvature -- The latter is TR-odd and is directly involved in the noncanonical dynamical structure of semiclassical Bloch electrons~\cite{Xiao2010}, while $\mathcal{\omega}_{n}\left(\mathbf{k}\right)$ results from electric field-induced Fermi surface shift.

It is also interesting to note that $\mathcal{\omega}_{n}\left(
\mathbf{k}\right)  =-\operatorname{Im}\mathcal{C}$, where $\mathcal{C}%
=\sum_{n_{1}\neq n}(\mathbf{A}_{nn_{1}}\times\mathbf{v}_{n_{1}n}%
^{\text{sys}})_{z}$. Being the imaginary part of $\mathcal{C}$, the $k$-space vorticity
of the layer current is connected to the real-space circulation of this
current around the electron wave-packet center $\mathbf{r}_{c}%
$~\cite{Xiao2010}: $\langle(\mathbf{\hat{r}}-\mathbf{r}_{c})\times
\hat{\mathbf{v}}^{\text{sys}}\rangle=\operatorname{Re}\mathcal{C}$. While
$\operatorname{Re}\mathcal{C}$ stems from the self-rotational motion of the
electron wave packet, $\omega_{n}(\mathbf{k})=-\operatorname{Im}%
\mathcal{C}$ results from the center-of-mass motion. Their relation is
analogous to the $k$-space Berry curvature and quantum metric, which are the
imaginary and real parts of the quantum geometric tensor~\cite{Ma2010}.

Nonzero layer current vorticity $\omega_{n}(\mathbf{k})$ and a net flux $\mathcal{V}$ require the quantum interlayer
hybridization of electronic states, which is a characteristic property not
shared by Berry curvature. First, in the absence of layer-hybridized states, $\omega_{n}(\mathbf{k})$ would vanish. Two scenarios of full layer polarization leading to vanishing $\omega_{n}(\mathbf{k})$ can be immediately identified: (i) If the states $|u_{n}\rangle$ and $|u_{n_{1}%
}\rangle$ involved in Eq.~(\ref{quantity}) are fully polarized in the same layer
around some $\mathbf{k}$, one gets $\mathbf{v}_{n_{1}n}^{\text{sys}}%
(\mathbf{k})\sim\mathbf{v}_{n_{1}n}(\mathbf{k})[1+l_{n}^{z}%
(\mathbf{k})]/2$ and thus
$\mathcal{\omega}_{n}\left(  \mathbf{k}\right)\sim0$. (ii) If
the two states are fully polarized in different layers, then $\mathbf{v}%
_{n_{1}n}^{\text{sys}}\left(  \mathbf{k}\right)\sim0$, and hence also $\mathcal{\omega}_{n}\left(
\mathbf{k}\right)\sim0$.
Moreover, by comparing the two forms of $\mathcal{V}$ in Eqs.~(\ref{FS}) and (\ref{sea}), one directly sees that a finite flux of vorticity also requires interlayer hybridization -- If $\ket{u_n}$ is fully layer-polarized, one has $\mathbf{v}_{n}^{\text{sys}}(\mathbf{k})\propto\mathbf{v}_{n}(\mathbf{k})$, so $\mathbf{v}_{n}(\mathbf{k})\times\mathbf{v}_{n}^{\text{sys}}(\mathbf{k})=0$ and $\mathcal{V}=0$ in Eq.~(\ref{FS}).

\begin{figure}[ptb]
	\centering
	\includegraphics[width=8.8cm]{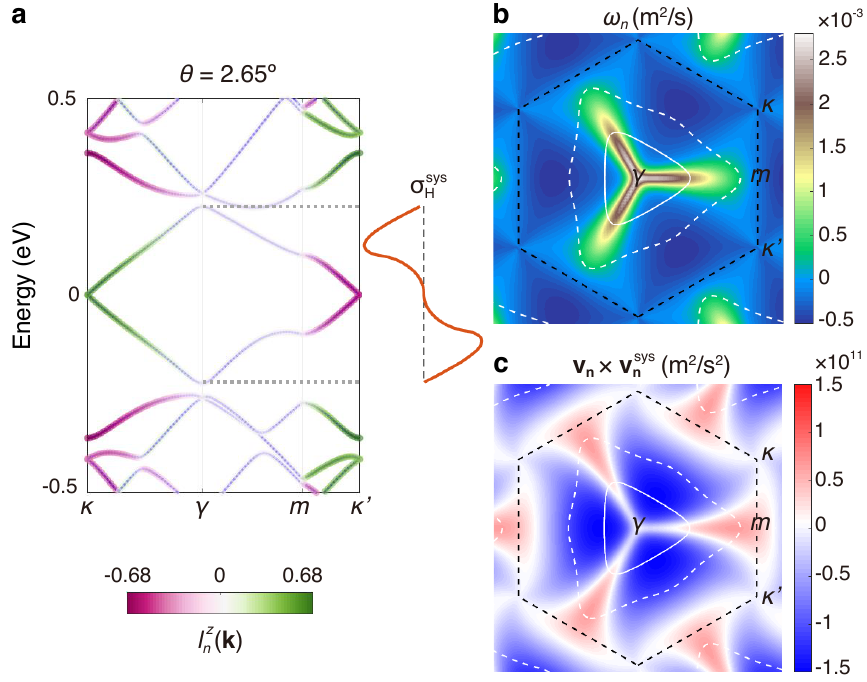}
	\caption{\textbf{Understanding the TR-even Hall effect from band geometric quantity.} \textbf{a} Band structure, \textbf{b} layer current vorticity $\omega_{n}$, and \textbf{c}
		$(\mathbf{v}_{n}\times\mathbf{v}_{n}^{\text{sys}})_{z}$ of the first conduction
		band from +K valley of $2.65^{\circ}$ tBG. In \textbf{a}, the color coding denotes
		the layer composition $l_{n}^{z}(\mathbf{k})$, and the brown curve shows the profile of central peaks of
		$\sigma_{\text{H}}^{\text{sys}}$. White curves in \textbf{b}, \textbf{c} show energy contours at $1/2$
		and $3/4$ of the band width. Black dashed hexagons in \textbf{b}, \textbf{c} denote the
		boundary of moir\'{e} Brillouin zone.}%
	\label{Fig:TBG_GeometryQuantities}%
\end{figure}


\textbf{Application to tBG.}
We now apply our theory to tBG. For
small twist angles $\theta$, we employ the continuum model with parameters
taken from Ref.~\cite{KoshinoTBGPRX2018}. The results are corroborated by
tight-binding calculations, which are also applicable at large $\theta$.
All model details are presented in the Methods section and Supplementary Note~2 and 3.

The calculation results for $\theta=1.05^{\circ}$ and $1.47^{\circ}$ are shown
in Figs.~\ref{Fig:TBG_theta}a and b. The central bands around zero energy
are separated from their neighbors with a global gap at such small angles.
When the Fermi level intersects the central bands, $\sigma_{\text{H}}^{\text{sys}}$
shows two narrow peaks with opposite signs for electron and hole doping. When
the Fermi level is located in the global gap, $\sigma_{\text{H}}^{\text{sys}}$ vanishes
as a Fermi surface property. Its magnitude starts to increase again when the
Fermi level is shifted outside the gap. Assuming a relaxation time of 1
ps~\cite{Dean2013,Schmitz2017,Sun2021}, the TR-even Hall conductivity can reach
dozens of $e^{2}/h$ upon Fermi level shifts within 20 meV. Such slight shifts
can be readily achieved by dual gates. The experimental measurement shall also
be facilitated by a large Hall ratio $\sigma_{\text{H}}^{\text{sys}}/\sigma_{xx}$,
which is independent of the relaxation time if the longitudinal charge conductivity
$\sigma_{xx}$ is also evaluated using the constant $\tau$ approximation. In
the current case, $\sigma_{xx}$ is strongly suppressed by the quite flat
dispersion, thus the Hall ratio can be $\gtrsim1$, as shown in the inset of
Figs.~\ref{Fig:TBG_theta}a and b.

The TR-even Hall effect is not restricted to long-wavelength moir\'{e} lattices. When $\theta$ gets large, the profiles of $\sigma_{\text{H}%
}^{\text{sys}}$ remain similar, but the width and magnitude of its peaks increase. This
is illustrated in Fig.~\ref{Fig:TBG_theta}c, where the central peaks of
$\sigma_{\text{H}}^{\text{sys}}$ are presented for a series of $\theta$. While the
two peaks become more separated as $\theta$ increases, sizable $\sigma
_{\text{H}}^{\text{sys}}\sim\mathcal{O}(1)\,e^{2}/h$ within dozens of meV around
zero-energy is still achievable for a wide range of $\theta$.

Next we look at the \textit{k}-space distributions of layer composition and layer current
vorticity to have a better understanding of the features of the TR-even Hall
effect. We illustrate these in Fig.~\ref{Fig:TBG_GeometryQuantities} using a
$2.65^{\circ}$ tBG and focus on the first conduction band. The band projection
of the layer composition $l_{n}^{z}(\mathbf{k})$ is denoted by color in Fig.~\ref{Fig:TBG_GeometryQuantities}a. At
low energies, the layer hybridization is weak, thus the states are dominantly
system/environment (or top/bottom) layer polarized around $\kappa$/$\kappa^{\prime}$. At higher
energies, Dirac cones from the two layers intersect and hybridize strongly
around the $\gamma$ point, rendering $l_{n}^{z}(\mathbf{k})\sim0$. Such layer
polarizations or hybridizations in different band regions are also manifested
in Fig.~\ref{Fig:TBG_GeometryQuantities}b for the distribution of the layer
current vorticity. It is concentrated along the path from $\gamma$ to $m$,
which is characterized by regions with strongly layer-hybridized and
nearly-degenerate bands, and is suppressed in the layer polarized regions
around $\kappa$ and $\kappa^{\prime}$. White curves in
Fig.~\ref{Fig:TBG_GeometryQuantities}b show two different Fermi surfaces. At
low electron doping, $\sigma_{\text{H}}^{\text{sys}}$ is contributed by the dark blue
area in Fig.~\ref{Fig:TBG_GeometryQuantities}b with $\omega_{n}<0$, thus it
is negative and increases in magnitude as the Fermi level is lifted towards
the middle of the band [see Eq.~(\ref{sea}) and brown curve in
Fig.~\ref{Fig:TBG_GeometryQuantities}a]. As the Fermi level is further
increased, regions with highly concentrated $\omega_{n}>0$ start to
contribute, hence the magnitude of $\sigma_{\text{H}}^{\text{sys}}$ drops. Evolution of
$\sigma_{\text{H}}^{\text{sys}}$ with the Fermi level can also be understood from the
distribution of $\mathbf{v}_{n}\times\mathbf{v}_{n}^{\text{sys}}$ shown in
Fig.~\ref{Fig:TBG_GeometryQuantities}c. Since it is dominantly negative in
the blue regions, according to Eq.~(\ref{FS}), $\sigma_{\text{H}}^{\text{sys}}$ shall
be negative and maximal when the Fermi level locates around the middle of the band.

In a long-wavelength moir\'e, the two valleys contribute to the layer Hall conductivity with identical sign and magnitude [see Eq. (7)], therefore intervalley scattering in tBG~\cite{ScatteringMinivalleyPRL2021} is not expected to diminish the effect. When the twist angle gets large enough, Umklapp process becomes important~\cite{MelePRB2010,HOTI_TBG_PRL2019}, and can hybridize the two valleys and lead to new features in the layer Hall conductivity that are not expected from the continuum model. Our tight-binding calculations for $\theta=21.8^\circ$ shows that
Umklapp process leads to emergence of new conductivity peaks at low energies (see Supplementary Fig.~3).
\color{black}


\begin{figure}[t]
	\includegraphics[width=6.8cm]{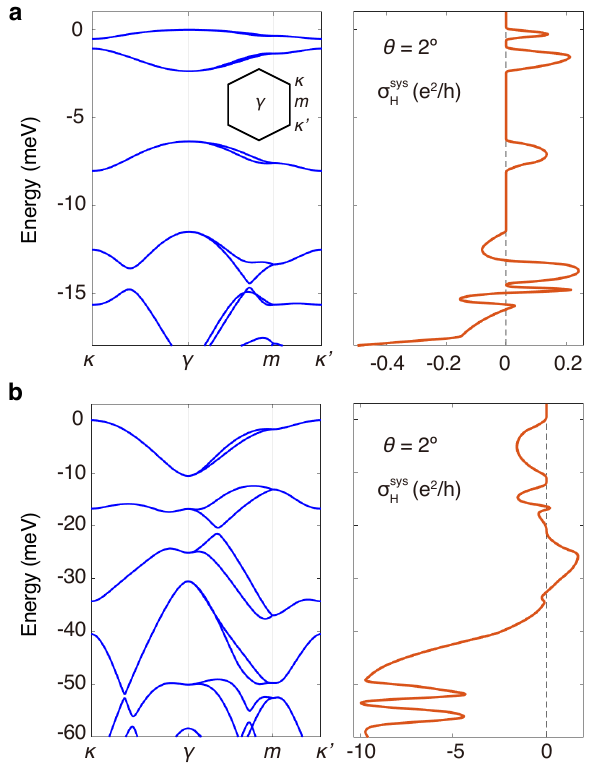} 
	\caption{\textbf{Valence band structure and TR-even Hall conductivity for tMoTe$_{2}$ at different $\theta$}. \textbf{a} $\theta=1.2^{\circ}$. \textbf{b} $\theta=3^{\circ}$.}%
	\label{Fig:tTMD_theta}%
\end{figure}

\begin{figure*}[t]
	\includegraphics[width=14cm]{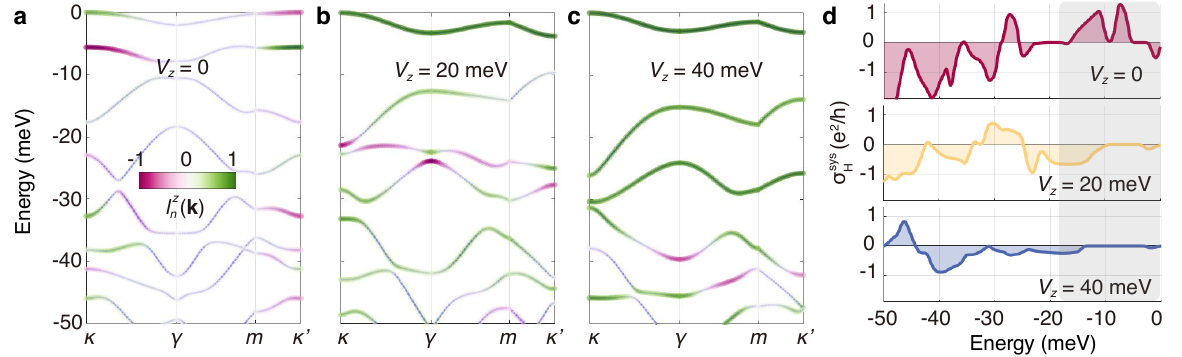} 
	\caption{\textbf{Gate control of the TR-even Hall effect in $2^{\circ}$ tMoTe$_{2}$}. \textbf{a}--\textbf{c} Valence bands from +K valley, and \textbf{d} TR-even Hall conductivities with different interlayer bias $V_{z}$. Color coding in \textbf{a}--\textbf{c} denotes the layer composition $l_{n}^{z}(\mathbf{k})$. The grey shaded area in \textbf{d} highlights the gradual suppression of the TR-even Hall effect by increasing $V_{z}$ at low energies.}%
	\label{Fig:tTMD_bias}%
\end{figure*}

\textbf{Application to tTMDs.}
Now we briefly address the TR-even
Hall effect in tTMDs and focus on its tuning. We consider the continuum model
of tMoTe$_{2}$~\cite{WuMacDonaldPRL2019,HongyiNSR} as an example (see Supplementary Note~1). The calculation results for $\theta=1.2^{\circ}$ and
$3^{\circ}$ are shown in Fig.~\ref{Fig:tTMD_theta}. The obtained TR-even Hall
conductivity can reach $\sim e^{2}/h$ and the Hall ratio $\sigma_{\text{H}%
}^{\text{sys}}/\sigma_{xx}\sim\mathcal{O}(0.1)$. It exhibits rich profiles, which can
be attributed to the complexity of tTMDs band structures that feature multiple
isolated narrow bands, and the efficient layer hybridization in this material.
As is shown in Fig.~\ref{Fig:tTMD_bias}a for the case of $\theta=2^{\circ}$,
most band regions are strongly layer hybridized.

Properties of tTMDs can be sensitively tuned via the interlayer bias $V_{z}$.
The band structures of $2^{\circ}$ tMoTe$_{2}$ with different $V_{z}$ are
shown in Figs.~\ref{Fig:tTMD_bias}a--c, and the corresponding TR-even Hall
conductivities are presented in Fig.~\ref{Fig:tTMD_bias}d. The profiles of
$\sigma_{\text{H}}^{\text{sys}}$ vary dramatically for different $V_{z}$, including the
magnitude and sign. A prominent impact of interlayer bias is the suppression
of $\sigma_{\text{H}}^{\text{sys}}$ at low energies (grey area in
Fig.~\ref{Fig:tTMD_bias}d), which indicates that the TR-even Hall effect can
be turned on/off with gate control. This is because the interlayer bias
polarizes low-energy electrons into one of the layers (see the dominance of green color in Figs.~\ref{Fig:tTMD_bias}b and c), thus reduces interlayer coupling.

This gate suppressed TR-even Hall effect totally differs from the gate induced
layer-polarized Hall effect appearing in the even-layer antiferromagnet
MnBi$_{2}$Te$_{4}$~\cite{Xu2021,Chen2022}. This distinction arises from the
fact that the former and latter effects favor strong layer hybridization and
layer polarization, respectively. It is also noted that the linear Hall effect
in a nonmagnetic bilayer can only appear in the layer-counterflow manner in
line with the Onsager relation (Fig.~\ref{Fig:Scheme}c), irrespective of the gate voltage;
while the layer-resolved Hall effects in top and bottom ferromagnetic layers of
antiferromagnetic bilayer MnBi$_{2}$Te$_{4}$ can be quite different when the
combined symmetry of TR and space inversion is broken by the gate.\\

\noindent{\bf{Discussion}}\\
In summary, we have discovered the TR-even linear charge Hall effect in a non-isolated 2D crystal endowed by the twisted interfacial coupling to an environmental layer, and elucidated the band origin of this effect. Measurable effects with great tunability are predicted in paradigmatic twisted bilayer materials in the presence of TR symmetry.
The layer Hall counterflow here contrasts with spin/valley Hall effect~\cite{Sinova2015,Xu2016valleytronics,Xiao2010}, in the latter counterflowing Hall currents for opposite spin/valley are spatially not separated so that it is impossible to access a charge Hall current. Here, the charge Hall current in each layer is experimentally accessible with the layer-contrasted geometry in vdW devices (see Supplementary Note~5).
It is also noted that the physics revealed is fundamentally distinct from the layer-polarized Hall effect in
layered antiferromagnetic insulators~\cite{Xu2021,Chen2022,Ma2022,Xie2022} and
the layer-dependent quantum Hall effect~\cite{BilayerHallGeimNatPhys2012,BilayerHallPhilipKimNatPhys2017,BilayerHallPabloNatNano2017,BilayerHallPhilipKimNatPhys2019,BilayerHallCoryDeanScience2022,BilayerHallPhilipKimPRL2017}%
, both of which rely on the TR symmetry breaking, and from the nonlinear Hall effects~\cite{Fu2015,Lai2021,LuHaizhouNatRevPhys2021} where the Onsager relation is obviously
irrelevant.
We note that in second-order nonlinear Hall effect, there also exists a TR-even band quantity -- Berry curvature dipole~\cite{LuHaizhouNatRevPhys2021}. It, however, has different symmetry constraints from the $k$-space vorticity of layer current. In particular, it is forbidden by the threefold rotational symmetry in 2D systems, and thus is absent in tBG and tTMDs studied here.
In the absence of magnetization and magnetic field, the sign of the linear Hall voltage now can be determined by the chirality of the interface, represented by the sign of the twisting angle. The effect therefore also serves as an efficient electrical probe on the structural chirality.\\


\noindent{\bf{Methods}}\\
\textbf{Continuum model of tBG and tTMDs.}
The top and bottom layers are rotated counterclockwise by $\pm\theta/2$, respectively,  with the corresponding rotation matrix $R_{\pm\frac{\theta}{2}}$. In tBG, the Hamiltonian for +K valley around $\bK_0=(\frac{4\pi}{3a},0)$ reads~\cite{KoshinoTBGPRX2018}
\begin{equation}
H=
\begin{pmatrix}
-\hbar v_F (\bk-\bK_t)\cdot R_{\frac{\theta}{2}}\bs&\mathcal{U}\\\mathcal{U}^\dagger&-\hbar v_F (\bk-\bK_b)\cdot R_{-\frac{\theta}{2}}\bs
\end{pmatrix}
\end{equation}
where $a=2.46$ \AA, $v_F=0.8\times10^6$ m/s, and $\bs=(-s_x,s_y)$ with $s_x=\begin{pmatrix}
0&1\\1&0
\end{pmatrix}$ and $s_y=\begin{pmatrix}
0&-i\\i&0
\end{pmatrix}$ the Pauli matrices in sublattice space.
$\bK_t=R_{\frac{\theta}{2}}\bK_0$ and $\bK_b=R_{-\frac{\theta}{2}}\bK_0$ are Dirac points after rotation in the top and bottom layer, respectively.
The moir\'e modulated interlayer tunneling is modeled by $\mathcal{U}=U_1+U_2e^{-i\bG_1\cdot\br}+U_3e^{-i(\bG_1+\bG_2)\cdot\br}$, where
\begin{equation}
\begin{aligned}
U_1&=
\begin{pmatrix}
u_{AA}&u_{AB}\\u_{AB}&u_{AA}
\end{pmatrix}\\
U_2&=
\begin{pmatrix}
u_{AA}&u_{AB}e^{i\frac{2\pi}{3}}\\u_{AB}e^{-i\frac{2\pi}{3}}&u_{AA}
\end{pmatrix}\\
U_3&=
\begin{pmatrix}
u_{AA}&u_{AB}e^{-i\frac{2\pi}{3}}\\u_{AB}e^{i\frac{2\pi}{3}}&u_{AA}
\end{pmatrix}
\end{aligned}.
\end{equation}
$\bG_1=-(\frac{1}{\sqrt{3}},\,1)\frac{2\pi}{L}$ and $\bG_2=(\frac{2}{\sqrt{3}},\,0)\frac{2\pi}{L}$ are the moir\'e reciprocal lattice vectors, with $L=a/(2\sin\frac{\theta}{2})$ the moir\'e period. The interlayer tunneling constants are $u_{AA}=79.7$ meV and $u_{AB}=97.5$ meV~\cite{KoshinoTBGPRX2018}.
The Hamiltonian from the -K valley can be obtained from TR operation.

The continuum model of tTMDs is very similar to that of tBG, with additional electrostatic modulations in each layer~\cite{WuMacDonaldPRL2019,HongyiNSR}. We leave its details to the Supplementary Note~1.

\noindent{\bf{Tight-binding model of tBG.}}
To characterize the electronic structures and TR-even Hall effect of tBG, we also use a tight-binding model following Ref.~\cite{KoshinoPRB2013}. The Hamiltonian is given by
\begin{equation}
\mathcal{H}=\sum_{\langle i, j\rangle} t (\bd_{ij})c_{i}^{\dagger} c_{j},
\end{equation}
where $c_{i}^{\dagger}$ and  $c_{j}$ are the creation and annihilation operators for the orbital on site $i$ and $j$, $\bd_{ij}$ represents the position vector from site $i$ to $j$, and $t (\bd_{ij})$ is the hopping amplitude between sites $i$ and $j$. We adopt the following approximations
\begin{equation}
\begin{aligned}
t(\bd) &=V_{p p \pi}\left[1-\left(\frac{\bd \cdot \mathbf{e}_{z}}{d}\right)^{2}\right]+V_{p p \sigma}\left(\frac{\bd \cdot \mathbf{e}_{z}}{d}\right)^{2} \\
V_{p p \pi} &=V_{p p \pi}^{0} \exp \left(-\frac{d-a_{0}}{\delta_{0}}\right) \\
V_{p p \sigma} &=V_{p p \sigma}^{0} \exp \left(-\frac{d-d_{0}}{\delta_{0}}\right)
\end{aligned}.
\end{equation}
In the above, $a_{0}\approx 1.42$~\AA~is the nearest-neighbor distance on monolayer graphene, $d_{0} \approx 3.35$~\AA~is the interlayer spacing, $V_{pp\pi}^{0}$ is the intralayer hopping energy between nearest-neighbor sites, and $V_{pp\sigma}^{0}$ is that between vertically stacked atoms on the two layers. Here we take $V_{pp\pi}^{0}=-2.7$ eV, $V_{pp\sigma}^{0}=0.48$ eV, $\delta_0$ is the decay length of the hopping amplitude and is set to 0.45255~\AA. The hopping for $d>20$~\AA ~is exponentially small thus is neglected in our study.\\

\noindent{\bf{Acknowledgements}}\\
We thank Xu-Tao Zeng for helpful discussions and Chengxin Xiao for the assistance of preparing figure 1.
This work is supported by the Research Grant Council of Hong Kong (AoE/P-701/20, W.Y.; HKU SRFS2122-7S05, W.Y.),
and the Croucher Foundation (Croucher Senior Research Fellowship, W.Y.). W.Y. also acknowledges support by Tencent Foundation.

\bibliographystyle{apsrev4-1}
\bibliography{LHE_ref}

\clearpage

\title{Supplementary Information}
\maketitle
\onecolumngrid

\setcounter{subsection}{0}
\setcounter{figure}{0}
\setcounter{equation}{0}
\setcounter{table}{0}

\renewcommand{\thesubsection}{\normalsize{Supplementary Note \arabic{subsection}}}
\renewcommand{\thepage}{S\arabic{page}}
\setcounter{page}{1}
\renewcommand{\thefigure}{\arabic{figure}}
\renewcommand{\thetable}{\arabic{table}}

\tableofcontents


\subsection{Continuum model of near $0^\circ$ twisted homobilayer TMD\lowercase{s}}

We assume that the top and bottom layers are rotated counterclockwise by $\pm\theta/2$ respectively with the corresponding rotation matrix $R_{\pm\frac{\theta}{2}}$. The Hamiltonian reads
\begin{equation}
H=
\begin{pmatrix}
\hbar v_F (\bk-\bK_t)\cdot R_{\frac{\theta}{2}}(s_x,s_y)+\text{diag}(E_g,0)+\mathcal{V}_t&\mathcal{U}\\\mathcal{U}^\dagger&\hbar v_F (\bk-\bK_b)\cdot R_{-\frac{\theta}{2}}(s_x,s_y)+\text{diag}(E_g,0)+\mathcal{V}_b
\end{pmatrix}
\end{equation}
around $\bK_0=(\frac{4\pi}{3a},0)$ for spin up carriers. Note that here zero energy is set at valence band edge. In the following, we consider MoTe$_2$, and use the parameters $a=3.472$ \AA, $v_F=0.4\times10^6$ m/s, monolayer band gap $E_g=1.1$ eV~\cite{WuMacDonaldPRL2019}. The rest of the notations are consistent with those in tBG in the Methods. To incorporate effects of interlayer bias, one adds $V_z/2$ and $-V_z/2$ to the two diagonal blocks, respectively.

The electrostatic modulation in the diagonal terms of $H$ are given by $\mathcal{V}_{l=t,b}=\begin{pmatrix} V_l^c&0\\0&V_l^v \end{pmatrix}$
with
\begin{equation}
\begin{aligned}
V_t^{c}&=V_0^c\sum_{i=1}^{3} \cos\left(\bG_i\cdot\br+\alpha_{c}\right)\\
V_b^{c}&=V_0^c\sum_{i=1}^{3} \cos\left(\bG_i\cdot\br-\alpha_{c}\right)\\
V_t^{v}&=V_0^v\sum_{i=1}^{3} \cos\left(\bG_i\cdot\br+\alpha_{v}\right)\\
V_b^{v}&=V_0^v\sum_{i=1}^{3} \cos\left(\bG_i\cdot\br-\alpha_{v}\right)
\end{aligned},
\end{equation}
where $V_0^c=11.94$ meV, $V_0^v=16$ meV, $\alpha_c=87.9^\circ$, and $\alpha_c=89.6^\circ$ in the case of MoTe$_2$~\cite{WuMacDonaldPRL2019}. Same as the case of tBG in the Methods section, $\bG_1=-(1/\sqrt{3},\,1)2\pi/L$, $\bG_2=(2/\sqrt{3},\,0)2\pi/L$, and $\bG_3=-\bG_1-\bG_2$.

The interlayer tunneling terms in the off-diagonal terms of $H$ are given by
\begin{equation}
\mathcal{U}
=
\begin{pmatrix}
u_{cc}&u_{cv}\\u_{vc}&u_{vv}
\end{pmatrix}
+
\begin{pmatrix}
u_{cc}&u_{cv}e^{-i\frac{2\pi}{3}}\\u_{vc}e^{i\frac{2\pi}{3}}&u_{vv}
\end{pmatrix}
e^{-i\bG_1\cdot\br}+
\begin{pmatrix}
u_{cc}&u_{cv}e^{i\frac{2\pi}{3}}\\u_{vc}e^{-i\frac{2\pi}{3}}&u_{vv}
\end{pmatrix}
e^{-i(\bG_1+\bG_2)\cdot\br},
\end{equation}
where $u_{cc}=-2$ meV, $u_{vv}=-8.5$ meV, and $u_{cv}=u_{vc}=15.3$ meV in the case of MoTe$_2$~\cite{WuMacDonaldPRL2019}.

We assign the Hamiltonian in the above as the +K valley in the main text. The Hamiltonian from the -K valley can be obtained from TR operation.


\subsection{Tight-binding model results for \lowercase{t}BG}

\begin{figure}[h]
	\includegraphics[width=5.3in]{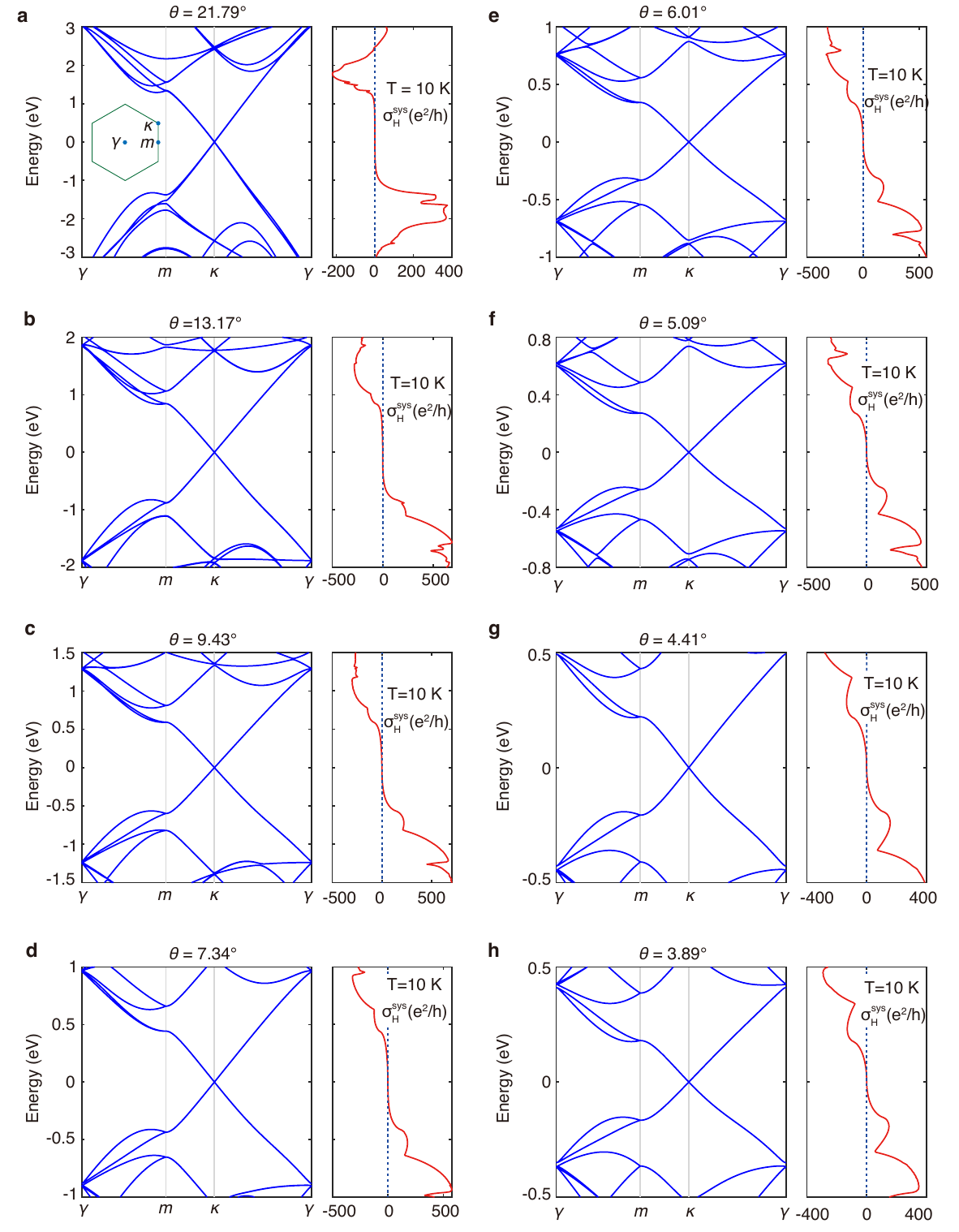}
	\renewcommand{\figurename}{\textbf{Supplementary Figure}}
	\caption{\textbf{Results from tight-binding calculations.} Band structures (left) and corresponding TR-even Hall conductivity (right) of tBG with \textbf{a} $\theta=21.79^{\circ}$, \textbf{b} $\theta=13.17^{\circ}$, \textbf{c} $\theta=9.43^{\circ}$, \textbf{d} $\theta=7.34^{\circ}$, \textbf{e} $\theta=6.01^{\circ}$, \textbf{f} $\theta=5.09^{\circ}$, \textbf{g} $\theta=4.41^{\circ}$ and \textbf{h} $\theta=3.89^{\circ}$. Note that spin degeneracy is taken into account in $\sigma_{H}^{\text{sys}}$.}
	\label{Fig:TBG_TBcollection}
\end{figure}

Supplementary Figure~\ref{Fig:TBG_TBcollection} shows the band structures and corresponding TR-even Hall conductivity in tBG from tight-binding calculations.

\subsection{Comparison of results from continuum and tight-binding models in \lowercase{t}BG}

\begin{figure}[h]
	\includegraphics[width=2.6in]{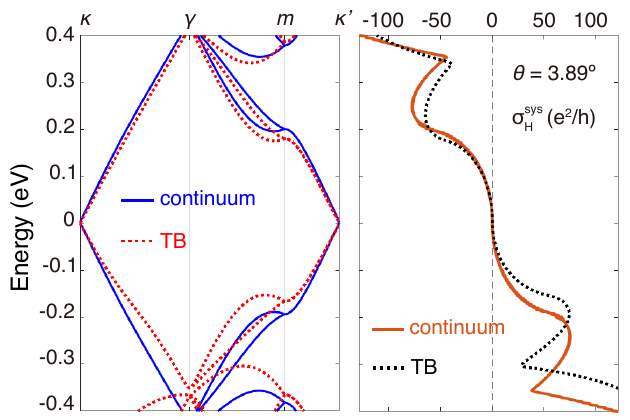}
	\renewcommand{\figurename}{\textbf{Supplementary Figure}}
	\caption{\textbf{Comparison of results from continuum model (solid curves) and tight-binding calculations (dashed curves).} (Left) Band structure, (right) TR-even Hall conductivity of $3.89^\circ$ tBG. Results of $\sigma_{H}^{\text{sys}}$ should be multiplied by a factor of 2 taken into account spin degeneracy.}
	\label{Fig:Continuum_vs_TB}
\end{figure}

Supplementary Figure~\ref{Fig:Continuum_vs_TB} shows the comparison of energy bands and TR-even Hall conductivity of $3.89^\circ$ tBG obtained from continuum model and tight-binding calculations. It is clear that the two methods yield consistent results. It should be noted that we have taken into account the effect of lattice relaxation in the continuum model by setting different values for interlayer tunneling between same-atom sites and different-atom sites~\cite{KoshinoTBGPRX2018}. While such effect is not included in tight-binding calculations. Better agreement could be achieved if such effect is considered or neglected~\cite{KoshinoPRB2013} in both methods.

\subsection{Effects of Umklapp intervalley process in $21.8^\circ$ \lowercase{t}BG}

\begin{figure}[h]
	\includegraphics[width=4in]{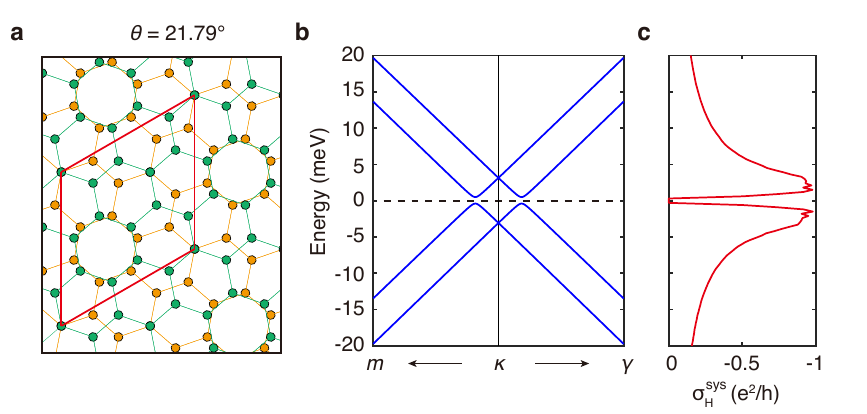}
	\renewcommand{\figurename}{\textbf{Supplementary Figure}}
	\caption{\textbf{a} Moir\'e of $\theta=21.8^\circ$ tBG. The red lines enclose one unit cell. \textbf{b} Low-energy band structures near the Dirac points. \textbf{c} Hall conductivity in the system layer.}
	\label{Fig:TBG_LHE21}
\end{figure}

Umklapp process becomes prominent near the Dirac points at large commensurate twist angles (e.g., $\theta=21.8^\circ$)~\cite{MelePRB2010,HOTI_TBG_PRL2019}. To examine the effect of Umklapp process, we have performed the tight-binding calculation for tBG at $\theta=21.8^\circ$ (Supplementary Figure~\ref{Fig:TBG_LHE21}). The obtained energy spectrum shows gap opening at the Dirac points (Supplementary Figure~\ref{Fig:TBG_LHE21}b), which is consistent with previous results~\cite{KoshinoPRB2013,HOTI_TBG_PRL2019}. Remarkably, the Hall conductivity shows new features (Supplementary Figure~\ref{Fig:TBG_LHE21}c) that are not expected from the continuum model. Note that the trend found in the small $\theta$ regime is that the conductivity peaks move to higher energy with the increase of $\theta$ and have opposite signs in the conduction and valence bands (Figure~2c of main text). At $\theta=21.8^\circ$, however, new conductivity peaks emerge at low energies with the same sign in the conduction and valence bands (Supplementary Figure~\ref{Fig:TBG_LHE21}c). This can be attributed to the Umklapp process.

\begin{figure}[h]
	\includegraphics[width=3.5in]{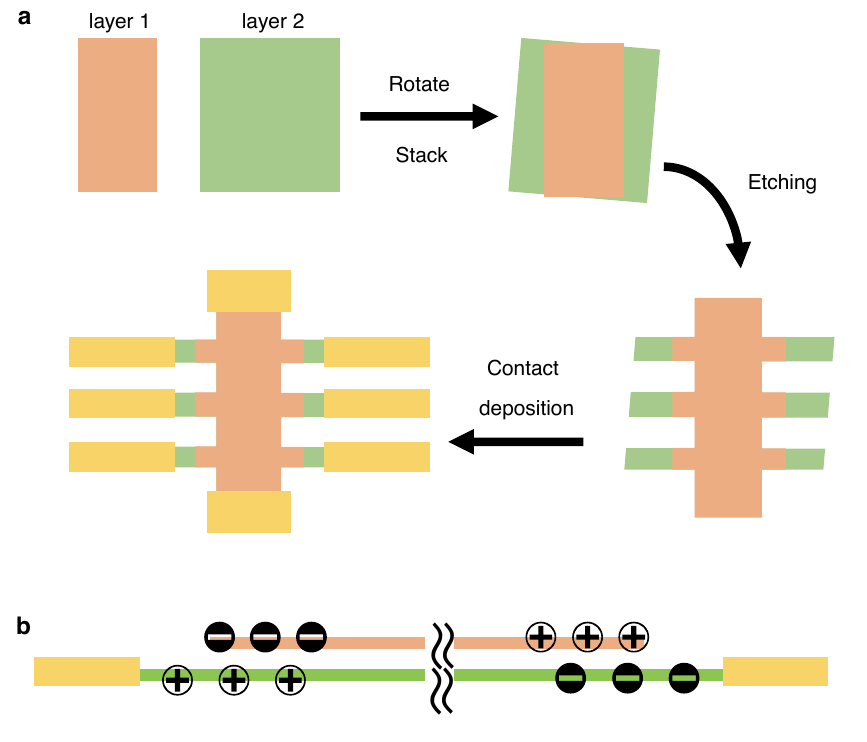}
	\renewcommand{\figurename}{\textbf{Supplementary Figure}}
	\caption{\textbf{a} Stacking two monolayers with different sizes/shapes (orange and green surfaces), and etching them into a Hall bar geometry. Electrical contacts (yellow) can be deposited onto the monolayer regions of Hall bars for layer-resolved measurement in \textbf{b}.}
	\label{Fig:Expt}
\end{figure}

\subsection{Proposal of experimental setup}

Fabrication of electrical contacts to one individual layer for layer-resolved measurement is experimentally feasible in coupled bilayer systems. Supplementary Figure~\ref{Fig:Expt}a shows the schematics of the experimental setup. When two monolayer flakes with different sizes/shapes are stacked, and etched into a Hall bar geometry, the overlapped part (twisted bilayer region) serves as the main channel connected to the source-drain contacts. Meanwhile, individual contacts can be deposited onto the monolayer parts of the Hall bar, which can be used as layer-resolved Hall voltage probes.

We notice that certain quantitative experimental uncertainties may exist, but they are not expected to qualitatively affect the observation. As a result of the predicted layer contrasted Hall effect, charges accumulate on the edges with layer and edge dependent signs (Supplementary Figure~\ref{Fig:Expt}b). The interface between the bilayer and monolayer regions on the Hall bar arms can have quantitative effect on the charge distribution in the measured layer (green layer in Supplementary Figure~\ref{Fig:Expt}), hence the transverse voltage drop measured by the electrical probes may deviate from the value predicted by our bulk theory without considering these details. However, order-of-magnitude reduction of the predicted effect is not expected when the contacts are close to the bilayer area.

\end{document}